\documentclass[twocolumn,nopacs,floatfix,amsmath,nofootinbib,amssymb,preprintnumbers]{revtex4-2}
\usepackage{graphicx,color,dcolumn,booktabs,bm,siunitx,multirow,mathrsfs}
\usepackage{txfonts}
\usepackage{xcolor}
\usepackage{ulem}
\usepackage{slashed}
\usepackage[colorlinks,
citecolor=blue,
anchorcolor=red,
menucolor=red,
linkcolor=red,
filecolor=red,
runcolor=red,
urlcolor=blue,
frenchlinks=false]{hyperref}

\definecolor{revisionpurple}{RGB}{128,0,128}

\begin{document}
	
	\title{Two-Pole Structure of $\Lambda(1405)$ with Temporal Evolution and Spatial Distribution}
	
	\author{Yu Zhuge}
	\author{Zhan-Wei Liu}\email{liuzhanwei@lzu.edu.cn}

	\affiliation
	{
		School of Physical Science and Technology, Lanzhou University, Lanzhou 730000, China\\
		Research Center for Hadron and CSR Physics, Lanzhou University and Institute of Modern Physics of CAS, Lanzhou 730000, China\\
		Lanzhou Center for Theoretical Physics, MoE Frontiers Science Center for Rare Isotopes, Key Laboratory of Quantum Theory and Applications of MoE, Key Laboratory of Theoretical Physics of Gansu Province, Gansu Provincial Research Center for Basic Disciplines of Quantum Physics, Lanzhou University, Lanzhou 730000, China
	}

	\begin{abstract}
		The $\Lambda(1405)$ is a special hadron resonance associated with two poles of the scattering amplitudes, and its nature remains under debate since its discovery before the birth of the quark model. In this work we study the structures of these two poles and their temporal evolution, including their difference, interference, and synergy. Each pole can usually be represented by the Gamow vector $|\psi^{\rm Gamow}\rangle$ in the complex momentum space $|\vec p e^{-i\theta}\rangle$. We construct its representation $|\psi^{\rm phys}\rangle$ in the real momentum space $|\vec p\rangle$ through the analytic continuation of the Gamow wavefunction, which also satisfies the Hamiltonian eigenvalue equation with the assistance of a virtual state vector. Both the decreasing behavior of the resonance and the production of the decayed scattering states can be simultaneously described by the temporal evolution $|\psi^{\rm phys},t\rangle=\exp(-iH t) \, |\psi^{\rm phys}\rangle$.
		The state $|\psi^{\rm phys},t=0\rangle$ gives the finite-range confinement of the resonance while $|\psi^{\rm phys},t\to \infty\rangle$ provides a Breit-Wigner-like distribution of the final scattering states whose appearance probability is nonzero as $r\to \infty$. In the two-channel system $\pi\Sigma$-$\bar{K}N$, we first dynamically generate the two poles of $\Lambda(1405)$ and then discuss their temporal evolutions and spatial distributions which can produce results consistent with experimental measurements such as the $\pi\Sigma$ invariant-mass spectrum and provide a new path to study hadron resonances.
		
	\end{abstract}

	\maketitle

	{\it Introduction} --- Since the 1950s, a large number of hadron resonances have been observed experimentally. Resonances are usually identified as peak structures in invariant-mass spectra or total cross sections in experiments, often described by Breit-Wigner type parametrizations~\cite{Breit:1936zzb}. Although such structures reflect the unstable nature of resonances, they are not equivalent to resonances themselves. In scattering theory, a resonance is naturally identified with a pole of the $S$-matrix on an unphysical Riemann sheet~\cite{Taylor:1972pty, Moiseyev:1998gjp,Willenbrock:2022smq}. However, the pole mass and width generally differ from the corresponding Breit-Wigner parameters, because the former are defined through the analytic structure of the amplitude in the complex-energy plane, whereas the latter characterize line shapes in the physical region. Establishing a natural connection between pole-related quantities and experimentally accessible observables therefore remains an active problem~\cite{Matsuyama:2006rp, Drechsel:2007if, Ronchen:2014cna, Briscoe:2023gmb, Doring:2025sgb,Baru:2003qq, Wang:2022vga, Wang:2023snv, Heuser:2024biq, Mannheim:2026eqq}.

	An unstable resonance necessarily evolves in time, and neglecting this fact may lead to conclusions that are difficult to interpret physically. For example, it is often stated that the wavefunction of a resonance grows exponentially as $r\to \infty$. Once time evolution is properly taken into account, this apparent divergence can be understood as a manifestation of the decay of the resonance into outgoing scattering states, which eventually propagate to infinity. The time evolution of unstable states has been investigated in Refs.~\cite{Fonda:1978dk, Chiu:1977ds, Khalfin:1981he, Norman:1988zz, Nakazato:1995cn, Rothe:2006rma, RamirezJimenez:2018jaj, Wang:2022kga, Yamada:2023wqp,Fock1962Criticism,Kelkar:2004zz, RamirezJimenez:2018dge,Garcia-Calderon_1992}. A resonance state can be described by the Gamow state~\cite{Gamow:1928zz}, and the related concept of compositeness was originally proposed by Weinberg as a criterion for determining whether a particle is composite or elementary~\cite{Weinberg:1965zz}. Recent developments for both bound states and resonances can be found in Refs.~\cite{Baru:2003qq, Kalashnikova:2009gt, Gamermann:2009uq, Baru:2010ww, Hanhart:2011jz, Hyodo:2011qc, Aceti:2012dd, Sekihara:2015gvw, Kamiya:2015aea, Guo:2015daa, Sekihara:2016xnq, Kang:2016ezb, Oller:2017alp, Li:2021cue, Wang:2021atf, Wang:2022vga, Kinugawa:2024crb}. In the framework developed in this work, the resonance itself is confined within a finite spatial region, while the outgoing components are generated dynamically through its time evolution.

	The $\Lambda(1405)$, located just below the $\bar{K}N$ threshold and observed mainly through the $\pi\Sigma$ channel, is one of the clearest systems where this distinction between poles and line shapes becomes essential. Its interpretation as a $\bar{K}N$ quasi-bound state or a dynamically generated meson-baryon resonance has since motivated extensive studies of the coupled $\pi\Sigma$ and $\bar{K}N$ dynamics~\cite{Hyodo:2011ur,Meissner:2020khl,Mai:2020ltx,Hyodo:2020czb,Liu:2016wxq,Menadue:2011pd,BaryonScatteringBaSc:2023ori,BaryonScatteringBaSc:2023zvt,Hyodo:2011qc,Aceti:2012dd,Wang:2021lth}. The $\Lambda(1405)$ region can contain two nearby isoscalar poles with chiral effective field theory~\cite{Kaiser:1995cy,Oset:1997it,Oller:2000fj,Jido:2003cb,Ikeda:2011pi,Ikeda:2012au, Lu:2022hwm, Molina:2015uqp, Nam:2017yeg, Ren:2021yxc, Xie:2023cej, Guo:2023wes, Zhuang:2024udv}, and recent amplitude analyses and direct $d(K^-,n)\pi\Sigma$ measurements further constrain the pole positions~\cite{Sadasivan:2022srs,J-PARCE31:2022plu}. The $\pi\Sigma$ line shape has been studied in experiments ~\cite{Hemingway:1984pz,Siebenson:2013rpa,Wickramaarachchi:2022mhi,BESIII:2024jgy, Wang:2026daa}, and these results show that the measured line shape is reaction dependent and should not be identified with a single Breit-Wigner resonance without considering the underlying pole structures.

	We start from the Gamow vector $|\psi^{\rm Gamow}\rangle$ in the complex momentum basis and analytically continue it to the real momentum Hilbert space, obtaining a physical component $|\psi^{\rm phys}\rangle$. With the assistance of a virtual component $|\psi^{\rm virtual}\rangle$, the combination $|\psi^{\rm phys}\rangle+|\psi^{\rm virtual}\rangle$ satisfies the Hamiltonian eigenvalue equation at the pole energy. The ordinary time evolution of $|\psi^{\rm phys}\rangle$ then separates the surviving resonance component from the produced scattering states. We analyze the interference and synergy of the two poles, which further discloses the structure of $\Lambda(1405)$ and reproduces the energy distribution of the asymptotic $\pi\Sigma$ and $\bar{K}N$ products that can be compared with experimental spectra.

	{\it The resonance wavefunction in complex momentum space} --- A resonance can be described by the Gamow vector $|\psi^{\rm Gamow}\rangle$, which is obtained by using the complex-scaling method with $\vec p\to \vec p_\theta\equiv \vec p e^{-i\theta}$ to solve the following Hamiltonian eigenvalue equation~\cite{Aguilar:1971ve,Balslev:1971vb} 
	\begin{align}\label{eq_CSM}
		H \, |\psi^{\rm Gamow}\rangle=(H_0+V) \, |\psi^{\rm Gamow}\rangle=E_{\rm pole} \, |\psi^{\rm Gamow}\rangle\, .
	\end{align}
	It is well known that this complex eigenmass $E_{\rm pole}=m_{\rm pole}-i \, \Gamma_{\rm pole} / 2$ is also the pole of the scattering $T$ matrix on the unphysical Riemann sheet where $T=V + VG_0T$ and $G_{(0)} = (E - H_{(0)} + i\epsilon)^{-1}$ \cite{Taylor:1972pty}. {In this work we consider the isoscalar S-wave $\pi\Sigma$-$\bar{K}N$ coupled-channel system relevant to the $\Lambda(1405)$, namely $|\alpha\rangle=|\pi\Sigma\rangle$ or $|\bar{K}N\rangle$.} Near a resonance pole, the $T$ matrix $T_{\alpha,\beta}(p, p'; E)\equiv \langle \alpha,p|T|\beta,p'\rangle$ can be approximated by
	\begin{align}\label{Texpand}
		T_{\alpha,\beta}(p, p'; E) \approx \frac{\gamma_{\alpha}(p) \,\gamma_{\beta}(p')}{E-E_{\rm pole}} +\cdots\, ,
	\end{align}
	and it can be shown that the Gamow wavefunction is related to the residue $\gamma_{\alpha}(p)$ of the $T$ matrix~\cite{Sekihara:2016xnq}.

	An alternative derivation of the relation between the Gamow wavefunction and the residue of the $T$ matrix can be obtained from the time-evolution operator. The transition amplitudes between the scattering channels can be simplified with $e^{-iHt} = (2\pi i)^{-1} \oint_{\mathcal C} dz \, e^{-izt} \, (z - H + i\epsilon)^{-1}$ and $G = G_0 + G_0 T G_0$:
	\begin{align}\label{residueEHIT}
		\langle\beta, {p}'| e^{-iHt} |\alpha, {p}\rangle = \frac{ \delta({p}'-{p})}{p^2} \delta_{\alpha,\beta} \, e^{-i\omega_\alpha(p)t} \qquad\qquad \nonumber\\
		\quad +\frac{1}{2\pi i} \oint_{\mathcal C} dz \, e^{-izt} \frac{T_{\beta,\alpha}(p', p;z)}{[z - \omega_\beta(p') + i\epsilon][z - \omega_\alpha(p) + i\epsilon]} \, ,
	\end{align}
	where $\omega_\alpha(p)$ is the kinetic energy of the channel $|\alpha\rangle$ and the contour $\mathcal C$ encloses the lower half of the complex $z$ plane. One can expand $ e^{-iHt} \, |\alpha, \vec{p}\rangle=C_{\alpha,p}(t) \, |\psi^{\rm Gamow}\rangle+|\mbox{other ordinary eigenstates} \rangle$ on the left side while the pole contribution can be extracted with the residue theorem, and then we obtain $C_{\alpha,p_\theta}(t)=\psi^{\rm Gamow}_\alpha(p_\theta) \, e^{-iE_{\rm pole}t}$ and
	\begin{align}\label{GamovRela}
		\psi^{\rm Gamow}_\alpha(p_\theta)\equiv\langle\alpha,p_\theta|\psi^{\rm Gamow}\rangle= \frac{\gamma_\alpha(p_\theta)}{E_{\rm pole} - \omega_\alpha(p_\theta)}\, ,
	\end{align}
	with the normalization convention used in Refs.~\cite{Hyodo:2011qc, Aceti:2012dd, Guo:2015daa, Sekihara:2016xnq, Oller:2017alp, Wang:2022vga, Kinugawa:2024crb}
	\begin{align}
		\sum_\alpha \int dp_\theta\, p_\theta^2\, \langle\alpha,p_\theta|\psi^{\rm Gamow}\rangle^2=1 \, .
	\end{align}

	{\it A natural transition to the representation in the real world} --- Although resonance states can be represented in the complex momentum basis, their direct physical interpretation requires a real-momentum representation, because measured momenta are real. We therefore introduce the physical representation $|\psi^{\rm phys}\rangle$, expanded in the real momentum basis.
	
	We first reproduce the two poles associated with the $\Lambda(1405)$ in the T matrix using a two-channel separable interaction in the $\pi\Sigma$ and $\bar{K}N$ channels
	\begin{align}
		V_{\alpha\beta}(p,p') &= \frac{3v_{\alpha\beta}}{4\pi^2f_\pi^2}\, u_\alpha(p) \, u_\beta(p')\, , \quad
		u_\alpha(p) = \left(1+\frac{p^2}{\Lambda_\alpha^2}\right)^{-2} \, .
	\end{align}
	By solving the Lippmann-Schwinger equation 
	\begin{align}\label{eq_BS}
		T_{\alpha,\beta}(p, p'; E) &= V_{\alpha\beta}(p,p') \notag\\
		&+ \sum_{\gamma} \int dk \, k^2 \frac{V_{\alpha,\gamma}(p,k) \,  T_{\gamma,\beta}(k, p';E)}{E - \omega_{\gamma}(k) + i \epsilon}  \, ,
	\end{align}
	we can search for poles on the Riemann sheet that is unphysical in the $\pi\Sigma$ channel ($\theta_{\pi\Sigma}\gg 0^\circ$) and physical in the $\bar{K}N$ channel ($\theta_{\bar K N}=0^\circ$). By adjusting parameters $f_\pi=92.4~{\rm MeV}$, $v_{\pi\Sigma,\pi\Sigma}=-0.4132$, $v_{\pi\Sigma,\bar{K}N}=0.2729$, $v_{\bar{K}N,\bar{K}N}=-0.4889$, $\Lambda_{\pi\Sigma}=0.6968~{\rm GeV}$, and $\Lambda_{\bar{K}N}=0.8055~{\rm GeV}$, we obtain the higher-mass (H) pole $E_{\rm pole}^{\rm H}=1430-22\,i~{\rm MeV}$ and the lower-mass (L) pole $E_{\rm pole}^{\rm L}=1338-89\,i~{\rm MeV}$.

	It is natural to analytically continue the Gamow wavefunction $\psi^{\rm Gamow}_\alpha(p_\theta)$ to the real momentum axis, from which we can define
	\begin{align}\label{eq_phys}
		|\psi^{\rm phys}\rangle = \sum_{\alpha}\int dp \, p^2\frac{\gamma_\alpha(p)}{E_{\rm pole} - \omega_\alpha(p)} \, |\alpha, {p}\rangle \, .
	\end{align}
	There are two vector states $|\psi^{\rm phys}_{\rm H}\rangle$ and $|\psi^{\rm phys}_{\rm L}\rangle$ corresponding to the two poles of $\Lambda(1405)$, and we omit the subscripts ``H/L'' for simplicity most of the time. One can reconstruct the wavefunction in the real momentum space from a finite basis expansion of complex-scaled wavefunctions, but such a back-rotation is an ill-posed inverse problem and may amplify numerical noise~\cite{Kruppa:2013ala}. Usually both the external momenta $p^{(\prime)}$ and the integrated momenta $k$ in Eq. (\ref{eq_BS}) are complex scaled to obtain $T_{\alpha,\beta}(p_\theta, p'_\theta; E\to E_{\rm pole})$. We only rotate the integrated momenta $k\to k_\theta$ and keep the external momenta $p^{(\prime)}$ in the real space to solve $T_{\alpha,\beta}(p, p'; E\to E_{\rm pole})$ and use the resulting residues to construct $|\psi^{\rm phys}\rangle$. Although the two wavefunctions $\psi^{\rm Gamow}_\alpha(p_\theta)$ and $\psi^{\rm phys}_\alpha(p)$ look very different, each contains the same pole information because they are connected by analytic continuation. Thus some nontrivial relations still exist, for example we notice
	\begin{align}
		&2 \, {\rm Re}\sum_\alpha\int dp_\theta \, p_\theta^2 \, \psi^{\rm Gamow*}_\alpha(p_\theta^*) \, \psi^{\rm Gamow}_\alpha(p_\theta) \nonumber \\
		=& \sum_\alpha\int dp\, p^2 \, \psi^{\rm phys*}_\alpha(p^*) \, \psi^{\rm phys}_\alpha(p) \, .
	\end{align}

	\begin{figure*}[tbp]
		\centering
		\includegraphics[width=\textwidth]{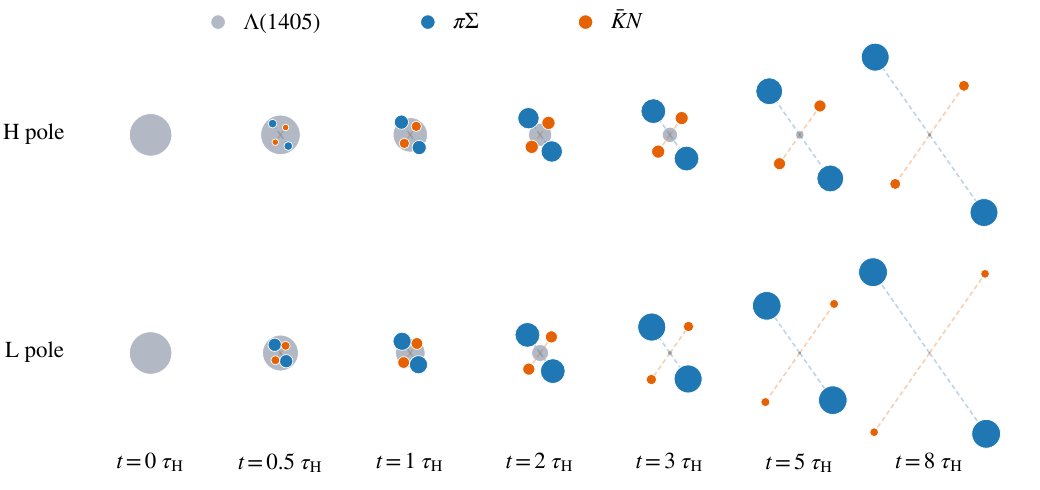}
		\caption{Illustration of the time evolution $|\psi^{\rm phys},t\rangle = e^{-iHt} \, |\psi^{\rm phys}\rangle$ of the resonance components. The gray area is proportional to the survival probability and decreases with time in units of $\tau_{\rm H}= 1 \, / \, |\Gamma_{\rm pole}^{\rm H}|$, while the produced $\pi\Sigma$ and $\bar{K}N$ scattering components move away from the interaction region. The distances of the scattering components from the center are proportional to their root-mean-square radii.}
		\label{evolution}
	\end{figure*}
	
	\begin{figure}[tbp]
		\centering
		\includegraphics[width=\columnwidth]{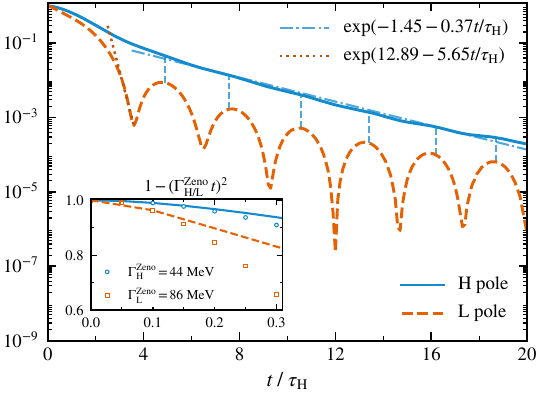}
		\caption{The survival probabilities of the two $\Lambda(1405)$ resonances, with time measured in units of $\tau_{\rm H}=1 \, / \, |\Gamma_{\rm pole}^{\rm H}|$. The blue solid and orange dashed curves are the numerical results for the two poles, respectively. The blue dot-dashed and brown dotted lines are the exponential fits to the higher-mass pole result over $3.5\leq t/\tau_{\rm H} \leq20$, and to the lower-mass pole result over $3.0\leq t/\tau_{\rm H} \leq3.6$. The blue vertical dashed segments mark phase-corresponding oscillatory regions of the two poles. The inset enlarges the short-time region $0\leq t/\tau_{\rm H} \leq0.3$. The open circles and squares show the quadratic forms $1-(\Gamma^{\rm Zeno}_{\rm H/L} \, t)^2$.}
		\label{psi0_time_evolution}
	\end{figure}
	
	Obviously $H \, |\psi^{\rm phys}\rangle\neq E_{\rm pole} \, |\psi^{\rm phys}\rangle$ because the eigenvalues of Hermitian operators are real in the ordinary Hilbert space. However, we can prove
	\begin{align}
		H \, (|\psi^{\rm phys}\rangle + |\psi^{\rm virtual}\rangle) = E_{\rm pole} \, (|\psi^{\rm phys}\rangle + |\psi^{\rm virtual}\rangle)\, 
	\end{align}
	by introducing a virtual vector state spanned around the basis states with the kinetic energy approaching the pole,
	\begin{align}
		|\psi^{\rm virtual}\rangle &=\sum_{\{\alpha \, | \, \theta_\alpha\neq 0\}}\oint_{\mathcal C_{\epsilon}^\alpha\to p_\alpha^{\rm on} } dp\, p^2 \frac{\gamma_\alpha(p)}{E_{\rm pole} - \omega_\alpha(p)} | \alpha,p\rangle,
	\end{align}
	where the contour $(\mathcal C_{\epsilon}^\alpha\to p_\alpha^{\rm on})$ refers to an infinitesimally small circle containing the point $p_\alpha^{\rm on}$ which satisfies $\omega_\alpha(p_\alpha^{\rm on})=E_{\rm pole}$. With respect to the Hamiltonian equation, $|\psi^{\rm Gamow}\rangle$ is equivalent to a vector $|\psi^{\rm phys}\rangle$ in the ordinary Hilbert space plus a virtual vector in the special complex space with the kinetic energies approaching the complex pole energy:
	\begin{align}
		|\psi^{\rm Gamow}\rangle \cong |\psi^{\rm phys}\rangle + |\psi^{\rm virtual}\rangle \, ,
	\end{align}
	which is a new essential relation between $|\psi^{\rm Gamow}\rangle$ and $|\psi^{\rm phys}\rangle$ in addition to their analytic connection.

	Although a resonance can be identified as a pole of the scattering amplitude, it is not a stationary state. It loses its resonance component and produces outgoing scattering states as time evolves, as illustrated in Fig.~\ref{evolution}. The produced scattering component cannot be directly described with $|\psi^{\rm Gamow}\rangle$. However, both parts can be naturally exhibited by the time evolution of $|\psi^{\rm phys}\rangle$
	\begin{align}\label{eq_psit}
		|\psi^{\rm phys},t\rangle = e^{-iHt} \, |\psi^{\rm phys}\rangle \, .
	\end{align}
	After inserting the completeness relation with $|\alpha,p\rangle$, we discretize the resulting momentum integrals using Gaussian quadrature and thus reduce the time evolution to a finite-dimensional matrix-vector operation for the numerical solutions.

	To describe both the remaining resonance and the produced scattering states, we expand 
	\begin{align}
		|\psi^{\rm phys},t\rangle = C(t) \, |\psi^{\rm phys}\rangle + |\chi_{\rm scatt.},t\rangle.
	\end{align}
	These two parts represent different states and should therefore be orthogonal, $\langle\psi^{\rm phys}|\chi_{\rm scatt.},t\rangle=0$, which also governs the conservation of probability. We stress that $|\psi^{\rm phys}\rangle$ is an ordinary vector in the familiar Hilbert space, so the expansion coefficient $C(t)$ and the wavefunction of $|\chi_{\rm scatt.},t\rangle$ can be obtained directly with the solution of $|\psi^{\rm phys},t\rangle$. The survival probability of the resonance state is then $|C(t)|^2$.

	The survival probabilities of the two pole states associated with the $\Lambda(1405)$  are shown in Fig.~\ref{psi0_time_evolution}, and one can see that they depart from a purely exponential law.  At very early times, the survival probability follows the expected quadratic behavior, known as the quantum Zeno effect~\cite{Wigner:1955zz, Wilkinson:1997sez, Chiu:1977ds, LEVITAN1988267, Nakazato:1995stb, Peshkin:2014jdw, RamirezJimenez:2018jaj, Crespi:2019smd}, and we insert a localized magnification and notice the survival probabilities in the short-time interval can be well described by $1-(\Gamma^{\rm Zeno}_{\rm H/L} \, t)^2$ with $\Gamma^{\rm Zeno}_{\rm H}=44$ MeV and $\Gamma^{\rm Zeno}_{\rm L}=86$ MeV. The fact that $\Gamma^{\rm Zeno}_{\rm H}\approx\Gamma_{\rm pole}^{\rm H}$ but $\Gamma^{\rm Zeno}_{\rm L} < \Gamma_{\rm pole}^{\rm L}$ indicates the complicated interference between these two nearby poles and the scattering channels. At late times for the lower-mass pole, its survival probability shows an obvious damped oscillatory behavior which is a known quantum phenomenon of resonance. In the intermediate-time region for the higher-mass pole, we show an exponential fit $\exp(-1.45-0.37t/\tau_{\rm H})$ over $3.5\leq t/\tau_{\rm H} \leq20$, and the deviation exhibits weak oscillations whose peaks marked by the vertical dashed lines coincide with the corresponding peaks in the survival probability of the lower-mass pole, which suggests these two states are strongly correlated.

	\begin{figure}[tbp]
		\centering
		\includegraphics[width=\columnwidth]{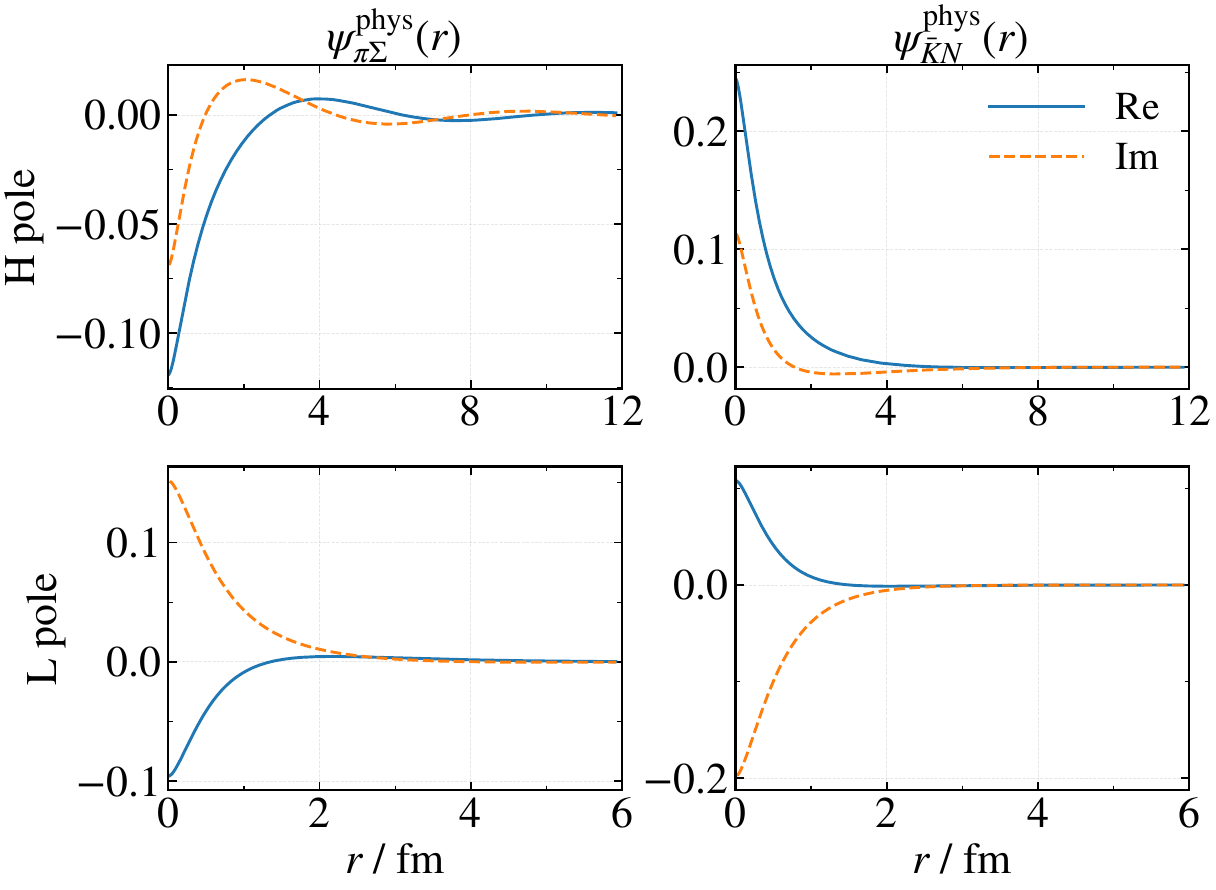}
		\caption{{The coordinate-space wavefunctions $\psi^{\rm phys}_\alpha(r)$ of the two $\Lambda(1405)$ poles in the real coordinate basis, in units of $\rm GeV^{3/2}$. The upper panels show the higher-mass pole and the lower panels show the lower-mass pole. The left and right columns show the $\pi\Sigma$ and $\bar{K}N$ channels, respectively.}}
		\label{physDisCord}
	\end{figure}

	The wavefunctions $\psi^{\rm phys}_\alpha(r)\equiv\langle\alpha,r|\psi^{\rm phys}\rangle$ and $\chi_{{\rm scatt.}\alpha}(r,t)\equiv\langle\alpha,r|\chi_{\rm scatt.},t\rangle$ in coordinate space can be obtained by Fourier transformation. {The wavefunctions of the two $\Lambda(1405)$ resonance components are plotted in Fig.~\ref{physDisCord}. The figure shows that they are localized in coordinate space and that the relative channel pattern differs between the higher-mass pole and the lower-mass pole.} 
	
	To visualize how the resonance decays into the final two-particle states and how these states propagate away from each other, we can calculate the root-mean-square radius $\sqrt{\bar{r}^2_{{\rm scatt.}\alpha}(t)}$ of each channel at time $t$, which is real in contrast to the generally complex radius associated with the Gamow state. {Fig.~\ref{evolution} illustrates the decay process of each resonance component into the $\pi\Sigma$ and $\bar{K}N$ final two-particle states. The area of each circle is proportional to the probability of the corresponding channel, while the separation between the meson and baryon is proportional to the root-mean-square radius. The decreasing gray circle shows the loss of the resonance component, and the colored circles show how the produced scattering states gradually move away from the interaction region.}

	{\it An application of the physical resonance representation to the experimental measurements} --- {The most relevant measurement for the $\Lambda(1405)$ is the final-state distribution in the $\pi\Sigma$ channel. In the present representation, one can first decompose each resonance component into its $\pi\Sigma$ and $\bar{K}N$ constituents through $\psi^{\rm phys}_\alpha(p)$, and then follow how these components evolve into physical scattering states through the coupled-channel $T$ matrix.}
	
	\begin{figure}[tbp]
		\centering
		\includegraphics[width=\columnwidth]{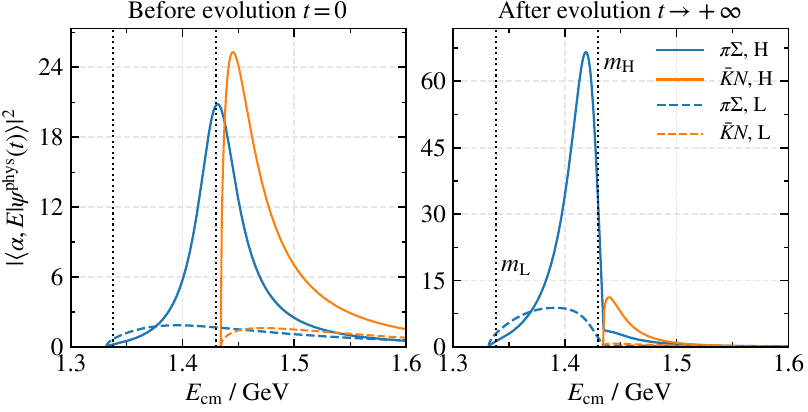}
		\caption{The energy distributions $|\langle \alpha,E \,| \,\psi^{\rm phys}(t) \rangle|^2$ for the $\pi\Sigma$ and $\bar{K}N$ channels. The left panel gives the distributions before evolution, $t=0$, and the right panel gives the asymptotic distributions after evolution, $t\to+\infty$. The black dotted lines denote the real part of each pole.}
		\label{WDis}
	\end{figure}
	
	At $t=0$, the energy distribution of each resonance component is given by $|\langle \alpha, E| \psi^{\rm phys} \rangle|^2$, as shown in the left panel of Fig.~\ref{WDis}. The figure provides a wavefunction of the initial states in the physical energy representation: the higher-mass pole has a pronounced strength around the $\bar{K}N$ threshold region, while the lower-mass pole is broader and more strongly connected to the lower $\pi\Sigma$ region. To obtain a distribution that can be compared with experiment, we evaluate the asymptotic state $|\psi^{\rm phys},t\to +\infty\rangle$ in the Schr\"odinger picture
	\begin{align}
		|\psi^{\rm phys},t\to +\infty  \rangle =\sum_{\alpha}\int dp \, p^2 \psi^{\rm phys}_\alpha(p)\times \qquad \qquad\qquad \nonumber\\
		\left(e^{-i\omega_\alpha(p)t}\,|\alpha, {p}\rangle+\sum_\beta\int dq \, q^2 \frac{T_{\beta,\alpha}(q, p;E)}{\omega_\beta(q) - \omega_\alpha(p) + i\epsilon} e^{-i\omega_\beta(q)t}\,|\beta, {q}\rangle \right)\, .
	\end{align}
	{The resulting distribution $|f_\alpha(E)|^2=|\langle \alpha,E \,|\, \psi^{\rm phys}(t=+\infty) \rangle|^2$ in the asymptotic limit is shown in the right column of Fig.~\ref{WDis}. Although the $\bar{K}N$ component is significant when the high-mass pole is initially formed, most of it turns into the $\pi\Sigma$ rather than $\bar{K}N$ scattering states because of energy conservation when $t\to \infty$ and the larger mass of $\bar{K}N$ than $m_{\rm pole}^{\rm H}$. The higher-mass pole contribution produces a relatively narrow $\pi\Sigma$ distribution peaked at around $1.419~{\rm GeV}$ in this model, while the lower-mass pole contribution gives a broader lower-mass distribution peaked at around $1.391~{\rm GeV}$, which shows that the peaks are much different from the pole masses.}

	\begin{figure}[tbp]
		\centering
		\includegraphics[width=\columnwidth]{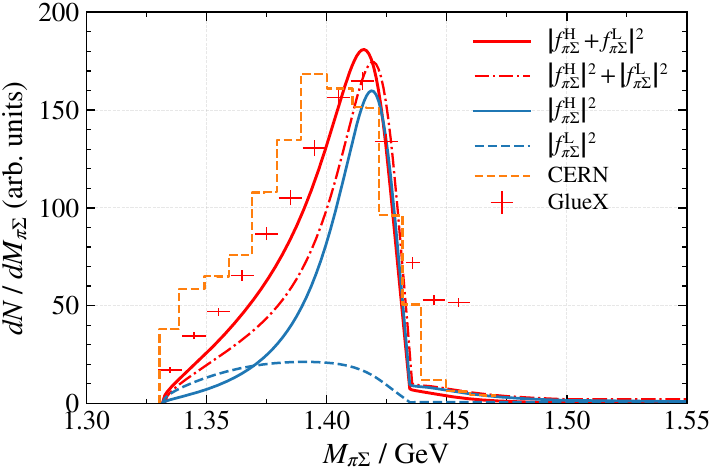}
		\caption{The $\pi\Sigma$ invariant-mass distribution. The higher-mass pole $|f_{\pi\Sigma}^{\rm H}(E)|^2$ and lower-mass pole curves $|f_{\pi\Sigma}^{\rm L}(E)|^2$ show the separate time-evolved contributions. The coherent sum $|f_{\pi\Sigma}^{\rm H}(E)+f_{\pi\Sigma}^{\rm L}(E)|^2$ is shown by the red solid line, and the sum results without any interference $|f_{\pi\Sigma}^{\rm H}(E)|^2+|f_{\pi\Sigma}^{\rm L}(E)|^2$ are shown by the red dash-dotted line. The CERN~\cite{Hemingway:1984pz} and GlueX~\cite{Wickramaarachchi:2022mhi} data sets are shown with arbitrary normalization for comparison of line shapes.}
		\label{PiSigmaInvMass}
	\end{figure}
	
	The $\pi\Sigma$ invariant-mass distribution obtained from the synergy of two poles is shown in Fig.~\ref{PiSigmaInvMass}. For each pole, we first use the time-evolved asymptotic state discussed above to obtain the outgoing $\pi\Sigma$ distribution $|f_{\pi\Sigma}^{\rm H/L}(E)|^2$. Then we give the full results with both the strongest interference scenario $|f_{\pi\Sigma}^{\rm H}(E)+f_{\pi\Sigma}^{\rm L}(E)|^2$ and no interference assumption $|f_{\pi\Sigma}^{\rm H}(E)|^2+|f_{\pi\Sigma}^{\rm L}(E)|^2$. From the figure, the currently observed $\pi\Sigma$ spectrum prefers the strongest interference scenario.

	{\it Summary} --- We have developed a physical description $|\psi^{\rm phys}_{\rm H/L}\rangle$ of the two-pole $\Lambda(1405)$ in the real momentum basis by
		analytically continuing the Gamow wavefunction from complex momentum space. The $|\psi^{\rm phys}_{\rm H/L}\rangle$ is spanned in the ordinary Hilbert space, and thus the probability interpretation can be naturally obtained. Moreover, the $|\psi^{\rm phys}_{\rm H/L}\rangle+|\psi^{\rm virtual}_{\rm H/L}\rangle$ also satisfies the Hamiltonian eigenvalue equation, as well as the Gamow wavefunction $|\psi^{\rm Gamow}_{\rm H/L}\rangle$.
	
	This physical representation allows the time evolution of each $\Lambda(1405)$ resonance component to be described in a transparent way. The evolved state can be decomposed into a surviving component and produced $\pi\Sigma$ and $\bar{K}N$ scattering states. The coordinate-space distributions and schematic time evolution show how the produced scattering states move away from the interaction region. The distribution at $t = 0$ tells us the channel content of each resonance component, whereas the asymptotic distribution at $t\to+\infty$ reflects the physical final-state spectrum. This provides a possible way to connect resonance wavefunctions defined through pole properties with physical distributions measured in scattering processes. The interference between the asymptotic distributions from the two poles associated with $\Lambda(1405)$ is essential for the $\pi\Sigma$ invariant-mass spectrum.

	\section*{ACKNOWLEDGMENTS}
	This work is supported by the National Natural Science Foundation of China under Grants No. 12335001, No. 12175091, No. 12247101, the ``111 Center'' under Grant No. B20063, and the innovation project for young science and technology talents of Lanzhou city under Grant No. 2023-QN-107.

	\bibliography{ref}

\end{document}